# Topological charge using cooling and the gradient flow

C. Alexandrou[a,b], A. Athenodorou[a,b], K. Jansen[c]

[a] Department of Physics, University of Cyprus, P.O. Box 20537, 1678 Nicosia, Cyprus
[b] Computation-based Science and Technology Research Center, The Cyprus Institute, 20 Kavafi Str., Nicosia 2121, Cyprus
[c] NIC, DESY, Platanenallee 6, D-15738 Zeuthen, Germany

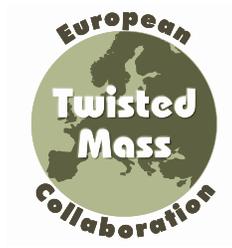

**Abstract**

The equivalence of cooling to the gradient flow when the cooling step $n_c$ and the continuous flow step of gradient flow $\tau$ are matched is generalized to gauge actions that include rectangular terms. By expanding the link variables up to subleading terms in perturbation theory, we relate $n_c$ and $\tau$ and show that the results for the topological charge become equivalent when rescaling $\tau \simeq n_c/(3 - 15c_1)$ where $c_1$ is the Symanzik coefficient multiplying the rectangular term. We, subsequently, apply cooling and the gradient flow using the Wilson, the Symanzik tree-level improved and the Iwasaki gauge actions to configurations produced with $N_f = 2 + 1 + 1$ twisted mass fermions. We compute the topological charge, its distribution and the correlators between cooling and gradient flow at three values of the lattice spacing demonstrating that the perturbative rescaling $\tau \simeq n_c/(3 - 15c_1)$ leads to equivalent results.

September 17, 2018

# Contents

# 1 Introduction

Besides the interest by itself, the calculation of the topological properties of gauge field configurations is needed for several investigations in lattice QCD. These may involve a direct use of the topological charge in observables or its use as a measure of auto-correlations. The former, for example, includes the computation of the $CP$-odd form factor $F_3$ and subsequently the neutron electric dipole moment (nEDM) [1]. This would shed light on the question whether the value of the nEDM is zero or not and can therefore give hints of possible beyond the standard model physics. There is a number of smoothing techniques that could be applied to extract the topological charge $\mathcal{Q}$, each one accompanied by its advantages and disadvantages [2, 3]. The gluonic definition of the topological charge density in Euclidean time-space is given by

$$q(x) = \frac{1}{32\pi^2}\epsilon_{\mu\nu\rho\sigma}\mathrm{Tr}\left\{G_{\mu\nu}G_{\rho\sigma}\right\}, \qquad (1)$$

with $G_{\mu\nu}$ the gluonic field strength tensor and $\epsilon_{\mu\nu\rho\sigma}$ the totally antisymmetric tensor. The introduction of the gradient flow [4, 5, 6] with its purturbatively proven renormalizability properties provides an attractive field-theoretic smoothing technique as compared to other techniques such as cooling and smearing, for which one can argue about the arbitrariness of their smoothing scale. The differential character of gradient-flow, however, makes it slower in comparison to other field-theoretic smoothers, such as cooling [7].

Recently it was demonstrated in Ref. [7] that using the Wilson action, gradient flow and cooling are equivalent if the gradient flow time $\tau$ and the number of cooling steps $n_c$ are appropriately matched. By expanding the link matrices perturbatively in the lattice spacing $a$ it was shown that at subleading order the two methods exhibit equivalence if one sets $\tau = n_c/3$. This analytic result was verified by a numerical investigation of a number of observables such as the average action and the topological susceptibility confirming that the two procedures indeed produce equivalent results. This suggests that in cases where high statistics are needed such as, for example, for the evaluation of higher moments of the topological charge [8], instead of using the more expensive gradient flow, one can opt to employing cooling to evaluate quantities of interest. Of course in some applications, such as the scale setting through $t_0$, where only a few hundreds of configurations are needed the computational cost is negligible and whether cooling or the gradient flow is used is not an important issue.

Studies that utilize dynamical quark simulations such as those pursued by the European Twisted Mass Collaboration (ETMC) [9, 10, 11] make use of configurations produced with Symanzik improved gauge actions, such as the Iwasaki and the Symanzik tree-level improved actions [12]. It is interesting to extend the study of Ref. [7] to explore the use of Symanzik improved actions in the smoothing procedure. This choice will alter the relation between the scales $\tau \simeq n_c/3$ since this depends on the choice of the smoothing action. We deliver the relation between gradient flow and cooling, by expanding the basic smoothing steps at subleading order in $a$ for Symanzik improved actions. Subsequently we test the validity of the formula numerically using ETMC configurations produced with $N_f = 2 + 1 + 1$ twisted mass fermions and the Iwasaki gauge action. In addition to the Wilson action, we employ as smoothing actions the Symanzik tree-level improved and the Iwasaki actions enabling us to generalize the correspondence. We test the equivalence on the topological charge itself as well as on the average action and the susceptibility. We also examine the degree of correlation among the results obtained with cooling and the gradient flow through the correlation coefficient. All observables suggest that the two smoothers become equivalent after a few transient cooling steps.

This article is organized as follows: In Section 2, we provide the relevant details regarding the production of the $N_f = 2 + 1 + 1$ configurations, in Section 3, we explain the different definitions of the topological charge density operators used for the calculation of the topological charge and in Section 4, we provide a short description of the cooling and gradient-flow techniques for smoothing a gauge configuration in order to set the ground for their analytical comparison. We then compare the two smoothers by expanding the link variables perturbatively in $a$. In Section 5 we provide numerical evidence of this equivalence by evaluating a number of relevant observables. Finally, in Section 6, we summarize and conclude.

## 2 Configurations

The gauge configurations are produced by the ETMC [9] using the Iwasaki improved action for the gluonic part

$$S_G = \frac{\beta}{N} \sum_x \left( c_0 \sum_{\substack{\mu,\nu=1 \\ 1 \leq \mu < \nu}}^{4} \{1 - \mathrm{ReTr}(U_{x,\mu,\nu}^{1\times 1})\} + c_1 \sum_{\substack{\mu,\nu=1 \\ \mu \neq \nu}}^{4} \{1 - \mathrm{ReTr}(U_{x,\mu,\nu}^{1\times 2})\} \right), \quad (2)$$

with $\beta = 2N/g_0^2$, $N = 3$ and $U_{x,\mu,\nu}^{1\times 1}$ the plaquette and $U_{x,\mu,\nu}^{1\times 2}$ rectangular ($1 \times 2$) Wilson loops. The Symanzik coefficients are set $c_0 = 3.648$ and $c_1 = -0.331$ and obey the relation $c_0 + 8c_1 = 1$. The twisted mass fermion action at maximal twist is employed. The formulation provides automatic $\mathcal{O}(a)$ improvement [13, 14], infrared regularization of small eigenvalues and fast simulations with dynamical fermions. For the doublet of light quarks the action is

$$S_F^{(l)}\left[\chi^{(l)}, \overline{\chi}^{(l)}, U\right] = a^4 \sum_x \overline{\chi}^{(l)}(x)(D_W[U] + m_{0,l} + i\mu_l \gamma_5 \tau^3)\chi^{(l)}(x), \quad (3)$$

where $\tau^3$ is the third Pauli matrix acting in the flavour space, $m_{0,l}$ the bare untwisted light quark mass and $\mu_l$ the bare twisted light quark mass. The massless Wilson-Dirac operator is given by

$$D_W[U] = \frac{1}{2}\gamma_\mu(\nabla_\mu + \nabla_\mu^*) - \frac{ar}{2}\nabla_\mu\nabla_\mu^*, \quad (4)$$

with the forward and backward covariant derivatives given by

$$\nabla_\mu \psi(x) = \frac{1}{a}\left[U_\mu(x)\psi(x+a\hat{\mu}) - \psi(x)\right] \quad \text{and} \quad \nabla_\mu^* \psi(x) = -\frac{1}{a}\left[U_\mu^\dagger(x-a\hat{\mu})\psi(x-a\hat{\mu}) - \psi(x)\right]. \quad (5)$$

The fields $\chi^{(l)}(x)$ are in the "twisted basis" and are related to the fields in the physical basis $\psi^{(l)}$ through the transformations

$$\psi^{(l)}(x) = \frac{1}{\sqrt{2}}\left(\mathbb{1} + i\tau^3\gamma_5\right)\chi^{(l)}(x) \quad \text{and} \quad \overline{\psi}^{(l)}(x) = \overline{\chi}^{(l)}(x)\frac{1}{\sqrt{2}}\left(\mathbb{1} + i\tau^3\gamma_5\right). \quad (6)$$

Apart from the doublet of light quarks, we also include a twisted heavy mass-split doublet $\chi^{(h)} = (\chi_c, \chi_s)$ for the strange and charm quarks. The associated action is expressed as

$$S_F^{(h)}\left[\chi^{(h)}, \overline{\chi}^{(h)}, U\right] = a^4 \sum_x \overline{\chi}^{(h)}(x)(D_W[U] + m_{0,h} + i\mu_\sigma \gamma_5 \tau^1 + \tau^3 \mu_\delta)\chi^{(h)}(x), \quad (7)$$

with $m_{0,h}$ the bare untwisted quark mass for the heavy doublet, $\mu_\sigma$ the bare twisted mass along the $\tau^1$ direction and $\mu_\delta$ the mass splitting in the $\tau^3$ direction. The heavy quark fields in the twisted basis are related to those in the physical basis through

$$\psi^{(h)}(x) = \frac{1}{\sqrt{2}}\left(\mathbb{1} + i\tau^1\gamma_5\right)\chi^{(h)}(x), \quad \overline{\psi}^{(h)}(x) = \overline{\chi}^{(h)}(x)\frac{1}{\sqrt{2}}\left(\mathbb{1} + i\tau^1\gamma_5\right). \quad (8)$$

Unless stated otherwise, the quark fields will be understood as "physical fields". The fermionic action in Eq. (3) breaks parity and isospin at non-vanishing lattice spacing with the latter inducing a cut-off effect of $\mathcal{O}(a^2)$ [14]. For more details on the twisted mass fermions see Ref. [9].

In order to test the equivalence between the two smoothing procedures we only need a single ensemble and a large number of configurations with a fine enough lattice spacing and relatively small pion mass. However, in order to investigate the behavior of observables as a function of the lattice spacing we include two additional ensembles, the pion mass of which is approximately the same as the one used for the more high statistics study. To this end, we selected the ensembles A60.24, B55.32 and D45.32sc in the notation of Ref. [15] at three different lattice spacings so the continuum limit can be taken. The details of the ensembles can be found in Table 1.

| A60.24, $\beta = 1.90$, $a = 0.094(1)$ fm, $r_0/a = 5.231(38)$ | | |
|---|---|---|
| $24^3 \times 48$, $L = 2.1$ fm | $a\mu$ | 0.0060 |
| | No. of confs | 1160 |
| | $a\, m_\pi$ | 0.17275(45)(23) |
| | $Lm_\pi$ | 4.15 |
| | $m_\pi$ | 0.362 GeV |
| B55.32, $\beta = 1.95$, $a = 0.082(1)$ fm, $r_0/a = 5.710(41)$ | | |
| $32^3 \times 64$, $L = 2.6$ fm | $a\mu$ | 0.0055 |
| | No. of confs | 4650 |
| | $a\, m_\pi$ | 0.15518(21)(33) |
| | $Lm_\pi$ | 4.97 |
| | $m_\pi$ | 0.372 GeV |
| D45.32sc, $\beta = 2.10$, $a = 0.064(1)$ fm, $r_0/a = 7.538(58)$ | | |
| $32^3 \times 64$, $L = 2.0$ fm | $a\mu$ | 0.0045 |
| | No. of confs | 949 |
| | $a\, m_\pi$ | 0.12087(40) |
| | $Lm_\pi$ | 3.89 |
| | $m_\pi$ | 0.368 GeV |

Table 1: Input parameters $(\beta, L, a\mu)$ of our lattice calculation for the ensembles A60.24, B55.32 and D45.32sc with the corresponding lattice spacing $a$, determined from the nucleon mass, and pion mass $am_\pi$ in lattice units.

## 3 Topological Charge

### 3.1 Definition of the Topological Charge on the Lattice

The topological charge of a gauge field is formally defined as the four–dimensional Euclidean integral over space-time

$$\mathcal{Q} = \int d^4x\, q(x)\,, \tag{9}$$

where the topological charge density $q(x)$ is defined in Eq. (1).

In practice, any valid lattice discretization of $q(x) \rightarrow q_L(x)$ leading to the right continuum expression of Eq. (1) can be used for the evaluation of the lattice equivalence of Eq. (9), given by

$$\mathcal{Q} = a^4 \sum_x q_L(x)\,. \tag{10}$$

However, depending on the discretization of the operator $q_L(x)$ lattice artifacts affecting the total topological charge $\mathcal{Q}$ vary. Hence, we do not expect to obtain an exact integer[1] value for the topological charge. Nevertheless, we expect that the total topological charge, for some definitions for the topological charge density, converge faster and are closer to an integer than that obtained by other definitions. To investigate the different definitions we use a number of lattice discretizations. The simplest lattice discretization, which can be constructed is based on the simple plaquette, depicted pictorially in Fig. 1

$$G_{\mu\nu}^{\text{plaq}}(x) = \text{Im} \left[ U_\mu(x) U_\nu(x + a\hat{\mu}) U_\mu^\dagger(x + a\hat{\nu}) U_\nu^\dagger(x) \right]\,, \tag{11}$$

---

[1]Of course one can obtain an exact integer when applying the Atiyah-Singer index theorem [16, 17] $Q = n_- - n_+$ and employing the number of Dirac zero modes $n_\pm$ with positive (+) and negative (-) chiralities obtained with the Overlap-Dirac operator [18].

with

$$q_L^{\text{plaq}}(x) = \frac{1}{32\pi^2}\epsilon_{\mu\nu\rho\sigma}\text{Tr}\left\{G_{\mu\nu}^{\text{plaq}}G_{\rho\sigma}^{\text{plaq}}\right\}. \tag{12}$$

This is a computationally cheap definition which, however, leads to lattice artifacts of order $\mathcal{O}(a^2)$. Nevertheless, this is still an adequate definition having been used in several determinations of the topological susceptibility in the past [19, 20].

Indubitably, the most common definition of the topological charge density is the clover definition given by

$$q_L^{\text{clov}}(x) = \frac{1}{32\pi^2}\epsilon_{\mu\nu\rho\sigma}\text{Tr}\left\{G_{\mu\nu}^{\text{clov}}G_{\rho\sigma}^{\text{clov}}\right\}, \tag{13}$$

with $G_{\mu\nu}^{\text{clov}}(x)$ the usual clover leaf (second picture in Fig. 1) defined as

$$\begin{aligned}
G_{\mu\nu}^{\text{clov}}(x) &= \frac{\text{Im}}{4}\Big[U_\mu(x)U_\nu(x+a\hat{\mu})U_\mu^\dagger(x+a\hat{\nu})U_\nu^\dagger(x)\\
&+ U_\nu(x)U_\mu^\dagger(x-a\hat{\mu}+a\hat{\nu})U_\nu^\dagger(x-a\hat{\mu})U_\mu(x-a\hat{\mu})\\
&+ U_\mu^\dagger(x-a\hat{\mu})U_\nu^\dagger(x-a\hat{\mu}-a\hat{\nu})U_\mu(x-a\hat{\mu}-a\hat{\nu})U_\nu(x-a\hat{\nu})\\
&+ U_\nu^\dagger(x-a\hat{\nu})U_\mu(x-a\hat{\nu})U_\nu(x+a\hat{\mu}-a\hat{\nu})U_\mu^\dagger(x)\Big].
\end{aligned} \tag{14}$$

However, this definition still carries a leading correction term of $\mathcal{O}(a^2)$. Hence, an improved definition of the topological charge density, which removes tree-level discretization errors and converges as $\mathcal{O}(a^4)$ in the continuum limit is also considered. Such a definition, given in Refs. [1, 21, 22], is

$$q_L^{\text{imp}}(x) = c_0 q_L^{\text{clov}}(x) + c_1 q_L^{\text{rect}}(x), \tag{15}$$

where $q_L^{\text{clov}}(x)$ is the ordinary clover topological charge density in Eq. (13) and $q_L^{\text{rect}}(x)$ is the clover-like operators where instead of squares we make use of horizontally– and vertically–oriented rectangular Wilson loops of size $2\times 1$ and $1\times 2$ respectively

$$q_L^{\text{rect}}(x) = \frac{2}{32\pi^2}\epsilon_{\mu\nu\rho\sigma}\text{Tr}\left\{G_{\mu\nu}^{\text{rect}}G_{\rho\sigma}^{\text{rect}}\right\}, \tag{16}$$

with

$$\begin{aligned}
G_{\mu\nu}^{\text{rect}}(x) &= \frac{\text{Im}}{8}\Big[U_\mu(x)U_\nu(x+a\hat{\mu})U_\nu(x+a\hat{\mu}+a\hat{\nu})U_\mu^\dagger(x+2a\hat{\nu})U_\nu^\dagger(x+a\hat{\nu})U_\nu^\dagger(x)\\
&+ U_\nu(x)U_\nu(x+a\hat{\nu})U_\mu^\dagger(x-a\hat{\mu}+2a\hat{\nu})U_\nu^\dagger(x-a\hat{\mu}+a\hat{\nu})U_\nu^\dagger(x-a\hat{\mu})U_\mu(x-a\hat{\mu})\\
&+ U_\mu^\dagger(x-a\hat{\mu})U_\nu^\dagger(x-a\hat{\mu}-a\hat{\nu})U_\nu^\dagger(x-a\hat{\mu}-2a\hat{\nu})U_\mu(x-a\hat{\mu}-2a\hat{\nu})U_\nu(x-2a\hat{\nu})U_\nu(x-a\hat{\nu})\\
&+ U_\nu^\dagger(x-a\hat{\nu})U_\nu^\dagger(x-2a\hat{\nu})U_\mu(x-2a\hat{\nu})U_\nu(x+a\hat{\mu}-2a\hat{\nu})U_\nu(x+a\hat{\mu}-a\hat{\nu})U_\mu^\dagger(x)\\
&+ U_\mu(x)U_\mu(x+a\hat{\mu})U_\nu(x+2a\hat{\mu})U_\mu^\dagger(x+a\hat{\nu}+a\hat{\mu})U_\mu^\dagger(x+a\hat{\nu})U_\nu^\dagger(x)\\
&+ U_\nu(x)U_\mu^\dagger(x-a\hat{\mu}+a\hat{\nu})U_\mu^\dagger(x-2a\hat{\mu}+a\hat{\nu})U_\nu^\dagger(x-2a\hat{\mu})U_\mu(x-2a\hat{\mu})U_\mu(x-a\hat{\mu})\\
&+ U_\mu^\dagger(x-a\hat{\mu})U_\mu^\dagger(x-2a\hat{\mu})U_\nu^\dagger(x-2a\hat{\mu}-a\hat{\nu})U_\mu(x-2a\hat{\mu}-a\hat{\nu})U_\mu(x-a\hat{\mu}-a\hat{\nu})U_\nu(x-a\hat{\nu})\\
&+ U_\nu^\dagger(x-a\hat{\nu})U_\mu(x-a\hat{\nu})U_\mu(x-a\hat{\nu}+a\hat{\mu})U_\nu(x+2a\hat{\mu}-a\hat{\nu})U_\mu^\dagger(x+a\hat{\mu})U_\mu^\dagger(x)\Big].
\end{aligned} \tag{17}$$

In order to remove the discretization error at tree-level one should use the Symanzik tree-level coefficients $c_1 = -1/12$ and $c_0 = 5/3$. A diagrammatic representation of the three definitions of $G_{\mu\nu}^r(x)$ ($r \equiv$ plaq, clov, rect) used in our investigation is provided in Fig. 1.

Ultraviolet fluctuations of the gauge fields entering in the definition of e.g. the topological charge lead to non-integer values. Thus, methods to suppress these ultraviolet fluctuations are employed. Such techniques include cooling and the more recently introduced gradient flow. We examine both these techniques using, beyond the Wilson, the Symanzik tree-level improved and Iwasaki actions.

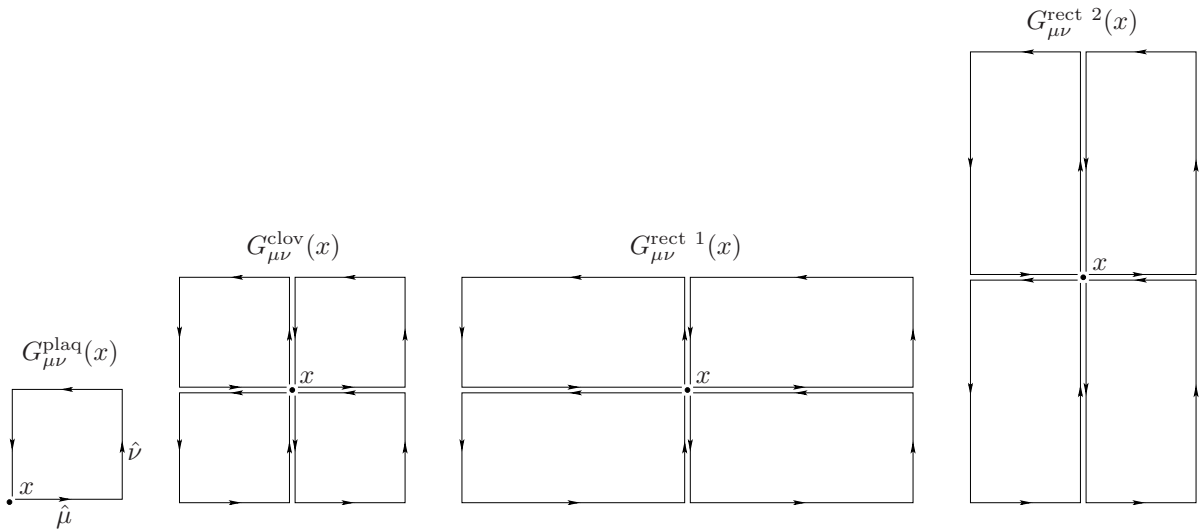

Figure 1: From left to right, we represent pictorially the plaquette operator used for the definition of the $G_{\mu\nu}^{\mathrm{plaq}}$, the ordinary clover $G_{\mu\nu}^{\mathrm{clov}}$ and the rectangle clovers $G_{\mu\nu}^{\mathrm{rect}\ 1}$, $G_{\mu\nu}^{\mathrm{rect}\ 2}$ such that $G_{\mu\nu}^{\mathrm{rect}} = G_{\mu\nu}^{\mathrm{rect}\ 1} + G_{\mu\nu}^{\mathrm{rect}\ 2}$.

## 4  Equivalence of Cooling with Gradient-Flow

We smooth out the ultraviolet fluctuations using the action in Eq. (2). The Symanzik coefficients must satisfy $c_0 + 8c_1 = 1$ and aside from this requirement, the value of $c_1$ can be chosen arbitrarily. The case of $c_1 = 0$ corresponds to the ordinary Wilson action. In addition to the Iwasaki action we also consider the Symanzik tree-level improved action with $c_1 = -1/12$. Any discrepencies resulting from different smoothing actions are interpeted as lattice artifacts and are expected to vanish in the continuum limit.

Smoothing a gauge link $U_\mu(x)$ can be accomplished by its replacement by some other link that minimizes the local action. To this purpose it makes more sense to rewrite the gauge action of Eq. (2) as

$$S_G = \frac{\beta}{N} \mathrm{ReTr}\{X_\mu^\dagger(x) U_\mu(x)\} + \{\text{terms independent of } U_\mu(x)\}, \qquad (18)$$

where $X_\mu(x)$ is the sum of all the path ordered products of link matrices, called the "staples", which interact with the link $U_\mu(x)$. The main components in the Wilson action are the plaquettes and thus the staples resulting from the square component of the action extend over $1 \times 1$ squares (in lattice units). For the rectangular part of the action the staples extend over rectangles of sizes $1 \times 2$ and $2 \times 1$. We can, therefore, write $X_\mu(x)$ as

$$X_\mu(x) = c_0 X_\mu^{\mathrm{plaq}}(x) + c_1 X_\mu^{\mathrm{rect}}(x), \qquad (19)$$

with

$$X_\mu^{\mathrm{plaq}}(x) = \sum_{\nu \geq 0, \nu \neq \mu} \left[ U_\nu(x) U_\mu(x + a\hat\nu) U_\nu^\dagger(x + a\hat\mu) + U_\nu^\dagger(x - a\hat\nu) U_\mu(x - a\hat\nu) U_\nu(x - a\hat\nu + a\hat\mu) \right], \qquad (20)$$

and
$$\begin{aligned}
X_\mu^{\text{rect}}(x) &= \sum_{\nu\geq 0,\nu\neq\mu}\Big[U_\nu(x)U_\nu(x+a\hat\nu)U_\mu(x+2a\hat\nu)U_\nu^\dagger(x+a\hat\nu+a\hat\mu)U_\nu^\dagger(x+a\hat\mu)+\\
&+ U_\nu^\dagger(x-a\hat\nu)U_\nu^\dagger(x-2a\hat\nu)U_\mu(x-2a\hat\nu)U_\nu(x-2a\hat\nu+a\hat\mu)U_\nu(x-a\hat\nu+a\hat\mu)\Big]\\
&+ \sum_{\nu\geq 0,\nu\neq\mu}\Big[U_\nu(x)U_\mu(x+a\hat\nu)U_\mu(x+a\hat\nu+a\hat\mu)U_\nu^\dagger(x+2a\hat\mu)U_\mu^\dagger(x+a\hat\mu)+\\
&+ U_\nu^\dagger(x-a\hat\nu)U_\mu(x-a\hat\nu)U_\mu(x-a\hat\nu+a\hat\mu)U_\nu(x-a\hat\nu+2a\hat\mu)U_\mu^\dagger(x+a\hat\mu)\Big]\\
&+ \sum_{\nu\geq 0,\nu\neq\mu}\Big[U_\mu^\dagger(x-a\hat\mu)U_\nu(x-a\hat\mu)U_\mu(x-a\hat\mu+a\hat\nu)U_\mu(x+a\hat\nu)U_\nu^\dagger(x+a\hat\mu)+\\
&+ U_\mu^\dagger(x-a\hat\mu)U_\nu^\dagger(x-a\hat\nu-a\hat\mu)U_\mu(x-a\hat\nu-a\hat\mu)U_\mu(x-a\hat\nu)U_\nu(x-a\hat\nu+a\hat\mu)\Big].
\end{aligned} \quad (21)$$

According to the above two equations, for a given link $U_\mu(x)$, the total number of plaquette and rectangular staples interacting with it is 6 and 18 respectively.

### 4.1 Cooling

Cooling is applied to a link variable $U_\mu(x)\in SU(N)$ by updating it, from an old value $U_\mu^{\text{old}}(x)$ to $U_\mu^{\text{new}}(x)$, according to the probability density

$$P(U)\propto \exp\left\{\lim_{\beta\to\infty}\beta\frac{1}{N}\text{ReTr}X_\mu^{\dagger}(x)U_\mu(x)\right\}. \quad (22)$$

The basic step of the cooling algorithm is to replace the given link $U_\mu^{\text{old}}(x)$ by an $SU(N)$ group element, which minimizes locally the action, while all the other links remain unaltered. This is done by choosing a matrix $U_\mu^{\text{new}}(x)\in SU(N)$ that maximizes

$$\text{ReTr}\{U_\mu^{\text{new}}(x)X_\mu^\dagger(x)\}. \quad (23)$$

In the case of an $SU(2)$ gauge theory, the maximization is achieved by

$$U_\mu^{\text{new}}(x)=\frac{X_\mu(x)}{\sqrt{\det X_\mu(x)}}. \quad (24)$$

For $SU(N)$ the maximization can be implemented by using the Cabibbo-Marinari algorithm [23]; one has to iterate the maximization over all the $SU(2)$ subgroups embedded into $SU(N)$.

We iterate this procedure so that all the links on all sites are updated. Such a sweep over the whole lattice is called a cooling step and will denote by $n_c$ the number of cooling steps performed. During the sweep the link variables, which have already been updated, are subsequently used for the update of the links still retaining their old value.

### 4.2 Gradient flow

The gradient flow is defined as the solution of the evolution equations [4, 5, 6]

$$\begin{aligned}
\dot V_\mu(x,\tau) &= -g_0^2\left[\partial_{x,\mu}S_G(V(\tau))\right]V_\mu(x,\tau)\\
V_\mu(x,0) &= U_\mu(x),
\end{aligned} \quad (25)$$

where $\tau$ is the total gradient flow time. In the above equation the link derivative is defined as

$$\begin{aligned}
\partial_{x,\mu}S_G(U) &= i\sum_a T^a\frac{\mathrm{d}}{\mathrm{d}s}S_G\left(e^{isY^a}U\right)\bigg|_{s=0}\\
&\equiv i\sum_a T^a\partial_{x,\mu}^{(a)}S_G(U),
\end{aligned} \quad (26)$$

with

$$Y^a(y,\nu) = \begin{cases} T^a & \text{if } (y,\nu) = (x,\mu) \\ 0 & \text{if } (y,\nu) \neq (x,\mu), \end{cases} \tag{27}$$

and $T^a$ ($a = 1, \cdots, N^2 - 1$) the Hermitian generators of the $SU(N)$ group. If we now set $\Omega_\mu = U_\mu(x)X_\mu^\dagger(x)$ we obtain

$$g_0^2 \partial_{x,\mu} S_G(U) = \frac{1}{2}\left(\Omega_\mu - \Omega_\mu^\dagger\right) - \frac{1}{2N}\text{Tr}\left(\Omega_\mu - \Omega_\mu^\dagger\right). \tag{28}$$

The last equation provides all we need in order to smooth the gauge fields according to the Eqs. (25). Evolving the gauge fields via gradient flow requires the numerical integration of Eqs. (25). This is performed using the third order Runge-Kutta scheme as explained in Ref. [6]. For the exponentiation of the Lie-algebra fields required for the integration, we apply the algorithm described in Ref. [24]. We investigate how the elementary integration step $\epsilon$ affects our results and find that $\epsilon = 0.02$ is a safe option as this was also pointed out in Ref. [7]; we observe that smaller elementary integration steps give the same results. We therefore set $\epsilon = 0.02$ for the integration step.

### 4.3 Perturbative relation between cooling and the gradient flow

Both cooling and gradient flow can be used to remove the ultraviolet fluctuations. Both should lead to the same topological properties provided that we are close enough to the continuum limit. Assuming that we are in the perturbative regime we can carry out a perturbative comparison in order to obtain an analytic relation between the scales involved in the two procedures following Ref. [7] where the relation

$$\tau \simeq n_c/3, \tag{29}$$

was derived for the Wilson action. In this work we derive a more general expression of the form $\tau = n_c \times f(c_1)$ for smoothing actions that, in addition to the plaquette, also include a rectangular term.

In the perturbative regime the link variables can be expanded as

$$U_\mu(x) \simeq \mathbb{1} + i\sum_a u_\mu^a(x)T^a, \tag{30}$$

with $u_\mu^a(x) \in \mathbf{R}$ is assumed to be infinitesimal.

Using Eqs. (20) and (21) the plaquette and rectangular staples are written as

$$X_\mu^{\text{plaq}}(x) \simeq 6 + i\sum_a w_\mu^a(x)T^a \quad \text{and} \quad X_\mu^{\text{rect}}(x) \simeq 18 + i\sum_a v_\mu^a(x)T^a, \tag{31}$$

where $w_\mu^a(x)$ and $v_\mu^a(x)$ are infinitesimal quantities. The leading coefficients with values 6 and 18 appearing in the above equations are just the number of plaquettes and rectangles interacting with the link on which the gradient flow evolution is applied. We can, therefore, write the sum of staples (Eq. (19)) as

$$X_\mu(x) \simeq 6c_0 + 18c_1 + ic_0 \sum_a w_\mu^a(x)T^a + ic_1 \sum_a v_\mu^a(x)T^a, \tag{32}$$

and, subsequently, $\Omega_\mu(x)$ as

$$\Omega_\mu(x) \simeq 6c_0 + 18c_1 + i\sum_a \left[(6c_0 + 18c_1)u_\mu^a(x) - \left(c_0 w_\mu^a(x) + c_1 v_\mu^a(x)\right)\right]T^a. \tag{33}$$

Hence, Eq. (28) becomes

$$g_0^2 \partial_{x,\mu} S_G(U) = i\sum_a \left[(6c_0 + 18c_1)u_\mu^a(x) - \left(c_0 w_\mu^a(x) + c_1 v_\mu^a(x)\right)\right]T^a. \tag{34}$$

Using the above expression, the evolution of the gradient flow can be approximated as

$$u_\mu^a(x, \tau + \epsilon) \simeq u_\mu^a(x, \tau) - \epsilon \left[ (6c_0 + 18c_1)u_\mu^a(x, \tau) - \left(c_0 w_\mu^a(x, \tau) + c_1 v_\mu^a(x, \tau)\right) \right]. \quad (35)$$

For the cooling procedure, one needs to consider that the link $U_\mu(x)$ is substituted with the projection of $X_\mu$ over the gauge group. Namely, for the case of an $SU(2)$ gauge theory this projection is manifested by Eq. (24) where we substitute $X_\mu(x)$ by Eq. (32). In the perturbative approximation this leads to [2]

$$U_\mu^{\text{new}}(x) \simeq \mathbb{1} + i \sum_a \frac{\left(c_0 w_\mu^a(x) + c_1 v_\mu^a(x)\right)}{6c_0 + 18c_1} T^a. \quad (38)$$

The above update corresponds to the substitution

$$u_\mu^a(x) \to \frac{\left(c_0 w_\mu^a(x) + c_1 v_\mu^a(x)\right)}{6c_0 + 18c_1}. \quad (39)$$

Comparing Eqs. [35] and [39] we observe that the gradient flow would evolve the same as cooling if one chooses a step of $\epsilon = 1/(6c_0 + 18c_1)$. In addition, during a whole cooling step the link variables, which have already been updated are subsequently used for the update of the remaining links that await update; this corresponds to a speed-up of a factor of two. Therefore, the predicted perturbative relation between the flow time $\tau$ and the number of cooling steps $n_c$ so that both smoothers have the same effect on the gauge field is

$$\tau \simeq \frac{n_c}{3c_0 + 9c_1} = \frac{n_c}{3 - 15c_1}. \quad (40)$$

The cooling/gradient flow rescaling factors for our choice of actions are given in Table 2.

| Smoothing action | $c_0$ | $c_1$ | $n_c/\tau$ |
|---|---|---|---|
| Wilson | 1 | 0 | 3 |
| Symanzik tr.level | $\frac{5}{3}$ | $-\frac{1}{12}$ | 4.25 |
| Iwasaki | 3.648 | -0.331 | 7.965 |

Table 2: Leading order perturbative rescaling between the number of cooling steps and gradient flow time such that the two smoothing techniques are equivalent. These numbers are according to Eq. (40).

An important question, which needs to be answered is how one tunes the smoothing parameters as the continuum limit is approached; this has been extensively discussed in Ref. [7] and we will briefly comment on how this is modified here. In practice, by applying the smoothing procedure on some configurations the ultraviolet (UV) properties of the theory up to some length scale $\lambda_S$ are modified by suppressing the UV fluctuations at smaller length scales. For this to be a viable procedure we need to show that by altering the UV part of the theory the continuum results remain unchangeable and, thus, the underlying physics does not depend on $\lambda_S$. Thus, one needs to choose the length scale $\lambda_S$, which for procedures like cooling often was taken arbitrarily; in other words the choice of the smoothing parameters such as $n_c$ in the case of cooling but

---

[2]This can be derived easily for $SU(2)$ where one can explicitly expand Eq. (24). Making use of the Mercator series expansion of the logarithm we write

$$\det X_\mu(x) = (6c_0 + 18c_1)^2 \det \left( \mathbb{1} + i \sum_a \frac{\left(c_0 w_\mu^a(x) + c_1 v_\mu^a(x)\right)}{6c_0 + 18c_1} T^a \right) = (6c_0 + 18c_1)^2 \left(1 + \mathcal{O}\left(\text{Tr}\{T^a\}\right) + \mathcal{O}\left(a^2\right)\right). \quad (36)$$

Thus, the expansion of Eq. (24) gives

$$\frac{X_\mu}{\sqrt{\det X_\mu(x)}} = \frac{1}{(6c_0 + 18c_1)} \left( 6c_0 + 18c_1 + ic_0 \sum_a w_\mu^a(x) T^a + ic_1 \sum_a v_\mu^a(x) T^a \right) = \mathbb{1} + i \sum_a \frac{\left(c_0 w_\mu^a(x) + c_1 v_\mu^a(x)\right)}{6c_0 + 18c_1} T^a. \quad (37)$$

also for other smearing techniques such as APE [25, 26], HYP [27, 28] and Stout [24, 29] is not entirely clear. The gradient flow, on the other hand, provides a smoothing procedure where this length scale is quantified as discussed below. Namely, it has been shown that one can simply renormalize composite operators at fixed physical flow time with

$$\lambda_S \simeq \sqrt{8t}, \tag{41}$$

with $t = a^2\tau$ being the gradient flow time in physical units. We can, therefore, translate the length scale $\lambda_S$ as a function of the cooling step $n_c$ according to the formula

$$\lambda_S \simeq a\sqrt{\frac{8n_c}{3 - 15c_1}}. \tag{42}$$

Given that the validity of Eq. (40) is confirmed numerically, we end up with an expression for an associate length scale $\lambda_S$ for the case of cooling as well. One can also generalize this correspondance for the cases of other smoothers, such as the APE and stout smearing [30].

As an example we consider the continuum limit of the topological susceptibility which is used in this work. According to Refs. [4, 6] one reads the topological susceptibility at a fixed value (in physical units) of $\lambda_S = \sqrt{8t} = O(0.1\text{fm})$ such that $\lambda_S$ is not too small so that discretisation effects are suppressed, as well as not too large so that the topological content of the gauge field is preserved. Practically, $\lambda_S$ should correspond to a plateau for the topological susceptibility which should be scale invariant. Hence, we extract the value of the topological susceptibility at fixed $\lambda_S$ for a sequence of lattice spacings and then extrapolate it in the continuum limit.

## 5 Numerical Results

### 5.1 Topological Charge

We apply cooling and gradient flow on $N_f = 2+1+1$ twisted mass fermions gauge configurations with $\beta = 1.90$, $\beta = 1.95$ and $\beta = 2.10$ using the Wilson (Eq. (2) with $c_1 = 0$), Symanzik tree-level improved (Eq. (2) with $c_1 = -1/12$) and Iwasaki (Eq. (2) with $c_1 = -0.331$) actions. We measure the average action, as well as the plaquette (Eq. (12)), clover (Eq. (13)) and improved (Eq. (15)) definitions of the topological charge for every cooling step $n_c$. Gradient flow is costlier and, thus, we take measurements for every $\Delta\tau = 0.1$ in units of gradient flow time (which corresponds to five integration steps for $\epsilon = 0.02$) instead of every integration step. We cover in total $60 - 100$ cooling steps while for the gradient flow we fix the maximum gradient flow time according to the perturbative expression of Eq. (40) and the maximum number of cooling steps. The cooling/gradient flow rescaling factors used are taken from Table 2.

The behavior of the topological charge $\mathcal{Q}$ for single configurations as a function of $n_c$ and $\tau$ is investigated for cooling and gradient flow, respectively for a given smoothing action and lattice spacing. In Fig. 2 we present the improved definition of the topological charge as a function of $n_c$ and $\tau$ for four randomly chosen configurations. We show results for $\beta = 1.90$ and $\beta = 2.10$. For $\beta = 1.90$ we observe that the topological charge for a given configuration whithin the whole range of $n_c \,/\, (3-15c_1) \times \tau$ yield different values for cooling and gradient flow. The difference in the value of the topological charge is not surprising since the different smoothers have different lattice artifacts and do not need to agree at non-zero values of the lattice spacing. For $\beta = 2.10$ the values become closer as expected. Thus, as one approaches the continuum limit the two different procedures converge. We note that the topological charge itself is not the main quantity of interest. It provides only a measure on the fluctuations and an input for the topological susceptibility, which is the physically relevant quantity. In the next section we will thus focus on the relevant physical observables. In this section, we restrict the presentation to the topological charge. Another observation from the results shown in Fig. 2 is that for the Wilson and Symanzik tree-level improved actions the topological charge $\mathcal{Q}$ as a function of $n_c$ or $\tau$ is not really constant. As can be seen in the left and middle panels of Fig. 2, the topological charge obtains different values with increasing $n_c$ and $\tau$. This behavior, although still present, appears to be supressed for our finest lattices with $\beta = 2.10$. Using the Iwasaki action, we observe that the topological charge fluctuates for $n_c \in [0, 20-30]$ and then becomes completely stable no matter what the lattice spacing is. These results have been observed when applying cooling in previous studies and they comply with theoretical expectations from an, admittedly, semi-classical picture. Namely, at finite lattice spacing the lattice action deviates from its continuum limit with deviations that increase as the gauge fields become larger. Instantons have a scale parameter $\lambda$, which enters non-trivially the action. As one decreases $\lambda$, the gauge fields are expected to become larger modifying the gauge action as well. The lattice action can be written [21, 31] (on dimensional grounds) as

$$S_{\text{Lat}}(a, \lambda) = S_{\text{cont}} \left\{ 1 + (a/\lambda)^2 \, a_2 + (a/\lambda)^4 \, a_4 + \mathcal{O}\left((a/\lambda)^6\right) \right\} \quad (43)$$

with $a_2 = -1/5$ for Wilson, $a_2 = 0, a_4 = -17/210$ for Symanzik and $a_2 = +2.972/5$ for Iwasaki. Stable instanton solutions require a lattice action which increases by decreasing the scale parameter $\lambda$. This requirement is fullfield only for the Iwasaki action and that is the reason why one observes stable topological charge. On the contrary, for the Wilson and Symanzik tree-level improved actions, the solutions are not stable; this is reflected in the fact that the values of the topological charge jump to different values. Nevertheless, stability sets in as $a \to 0$; this is visible for the case of $\beta = 2.10$ in Fig. 2 where we observe less changes in the value of $\mathcal{Q}$.

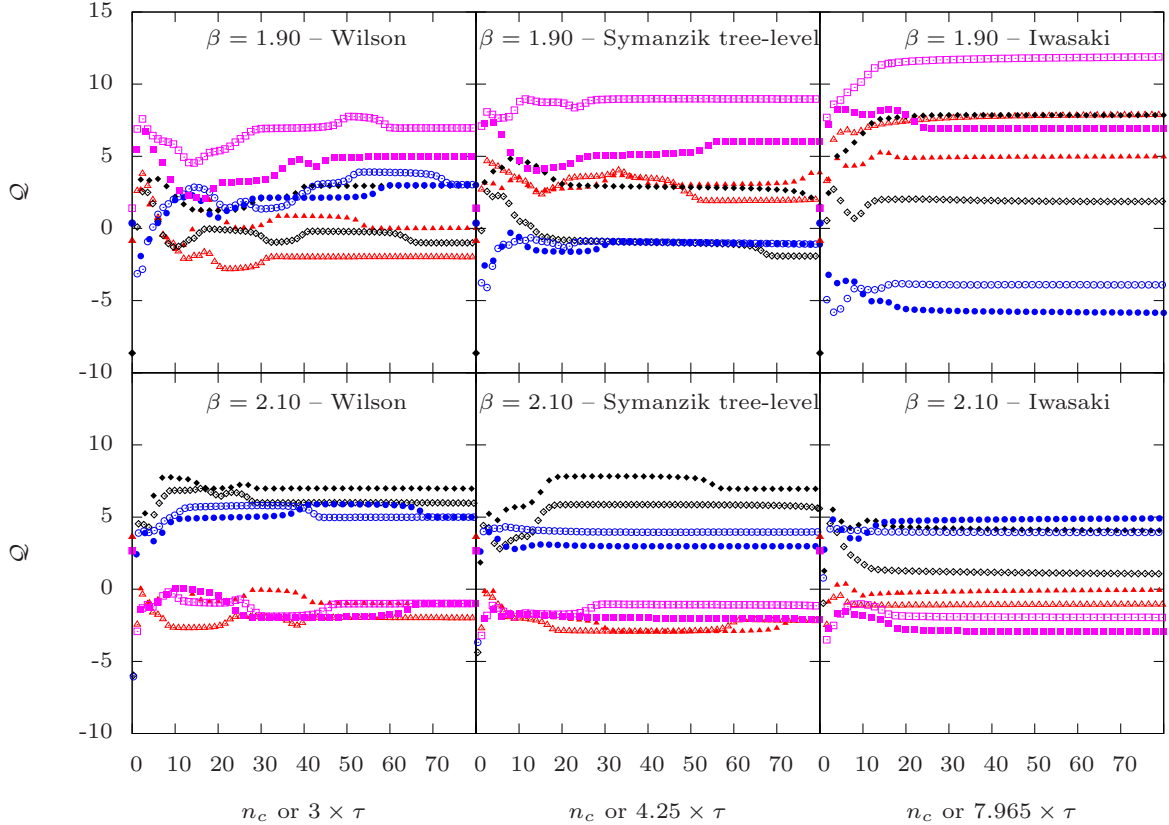

Figure 2: The improved definition of the topological charge as a function of $n_c$ for cooling and $\tau$ rescaled by a factor of 3, 4.25 and 7.965 for gradient flow extracted with Wilson (left), Symanzik tree-level improved (middle) and Iwasaki (right) smoothing actions respectively. The different colors correspond to the four different configurations chosen randomly while filled and open symbols correspond to cooling and gradient flow respectively. Upper panel is for $\beta = 1.90$ and lower panel for $\beta = 2.10$.

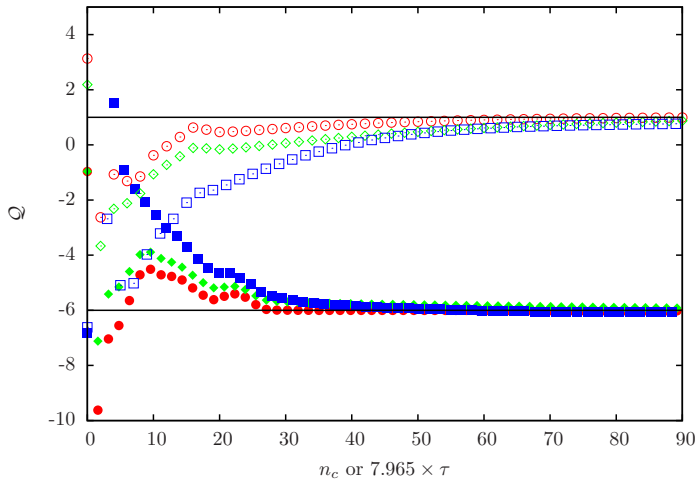

Figure 3: An example of the behavior of the topological charge for a single configuration as a function of $n_c$ and $\tau$ rescaled by 7.965 for cooling (open symbols) and gradient flow (filled symbols) for two different configurations. With the red circles we present the improved, with green diamonds the clover and with blue squares the plaquette definition of the topological charge. The smoothing has been performed with the Iwasaki action.

Comparing results for the three different definitions of the topological charge density, we observe that for the improved case the topological charge converges closer and faster to a near integer value compared to the other two definitions. All three definitions for the three ensembles give topological charges, which converge to the same near integer as a function of the relevant smoothing scale. These two observations suggest that indeed the three topological charge definitions differ only due to lattice artifacts. Such a comparison is meaningful only if the topological charge acquires stability and hence, we consider the Iwasaki action. In Fig. 3, we observe that for the clover as well as for the improved definition, the topological charge converges faster than when the plaquette definition is used in particular in the case when cooling is performed.

In Fig. 4 we present an example of the time history (first 900 configurations) of the topological charge $\mathcal{Q}$ for gauge configurations that have been cooled using $n_c = 50$. We also include the time history when using the gradient flow for a step of $\tau = n_c/(3 - 15c_1)$. Results are shown for $\beta = 1.95$ and $\beta = 2.10$ for the three gauge actions. As can be seen, the topological charge does not suffer from large autocorrelations and the time histories between cooling and the gradient flow are very similar. This similarity can be quantified by the calculation of the linear correlation coefficient, which is the topic of Section 5.4.

Additionally, in Fig. 5, we provide the histogram of the topological charge for both cooling and gradient flow for the three actions. We observe that the histograms exhibit nearly gaussian distributions in particular for the $\beta = 1.95$ ensemble where a large number of configurations is analyzed. As expected, the distributions using cooling or the gradient flow look very similar for all three actions and the associated gaussian fits fall on top of each other. This already points to the equivalence anticipated for the topological susceptibility.

## 5.2 Average Action Density

As a common scale for the two smoothing techniques we can use the action, the minimization of which defines both smoothers. Instead of looking at the action we consider the dimensionless average action density $\langle \bar{S}_\mathrm{G} \rangle$

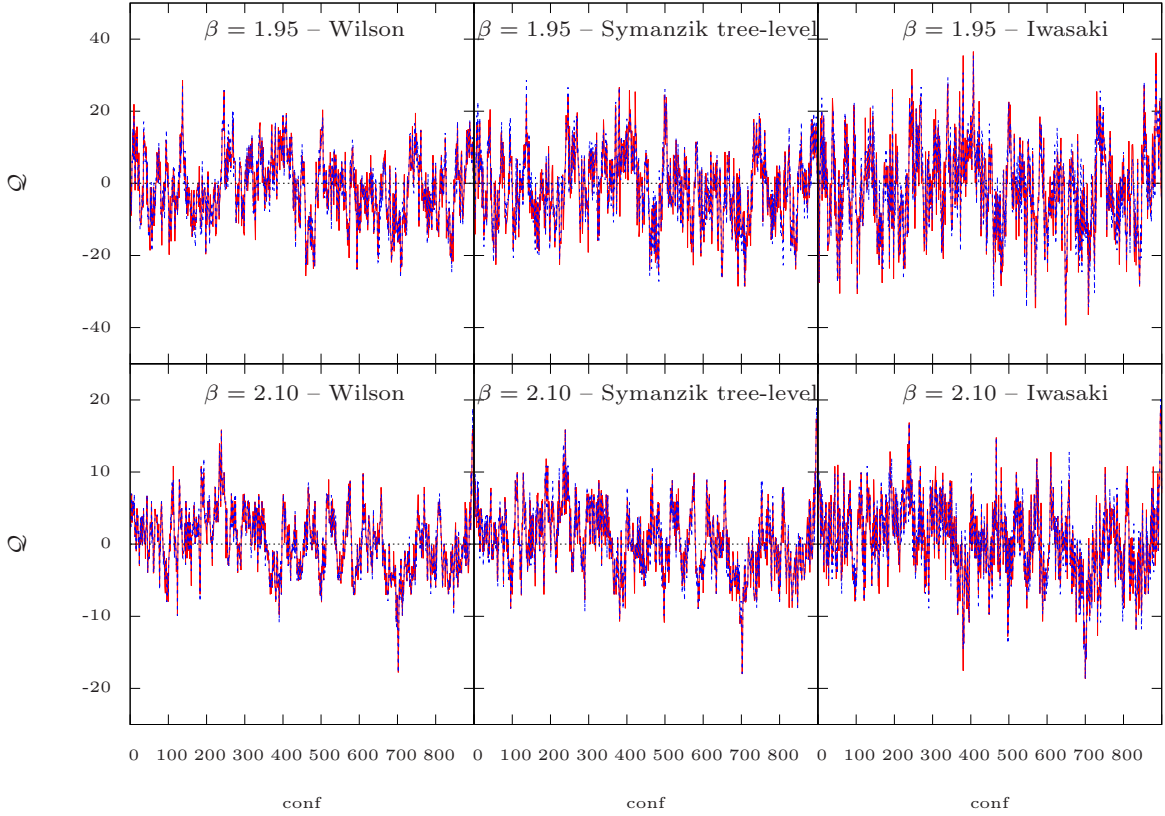

Figure 4: The time history of the topological charge which has been extracted by cooling (blue dashed line) and gradient flow (red solid line) at $n_c = 50$ and the corresponding flow time for each different choice of smoothing action. In the upper row we present results for $\beta = 1.95$ and in the lower results for $\beta = 2.10$.

$\in [0, 1)$ defined as

$$\langle \bar{S}_{\rm G} \rangle \;=\; 1 - \left\langle \frac{c_0 \sum_x \sum_{\substack{\mu,\nu=1\\ 1\leq\mu<\nu}}^{4} {\rm Re\,Tr}\, U^{1\times 1}_{x,\mu,\nu} + c_1 \sum_x \sum_{\substack{\mu,\nu=1\\ \mu\neq\nu}}^{4} {\rm Re\,Tr}(U^{1\times 2}_{x,\mu,\nu})}{(6c_0 + 12c_1)\, V a^{-4} N} \right\rangle. \qquad (44)$$

In Fig. 6 we present the average action density for $\beta = 1.95$ as a function of $n_c$ and the perturbatively determined values of the gradient flow time, namely $3\times$, $4.25\times$, and $7.965 \times \tau$ for the Wilson, the Symanzik tree-level improved and the Iwasaki actions, respectively. As expected from the findings of Ref. [7], for the Wilson action, the rescaling $n_c = 3\tau$ leads to equivalent results for this quantity between gradient flow and cooling for small values of $n_c$ and $\tau$. For instance for $\beta = 1.95$ where our results are more accurate we find that for $n_c \geq 20$ the average action for both procedures becomes the same. Our results show that a similar behavior is observed also for the other two actions. Namely, the average action density deviates for small values of the smoothing scales but for $n_c \sim 30$, for the Symanzik tree-level improved, and $n_c \sim 50$, for the Iwasaki action, they become equal. Similar behavior is also observed for $\beta = 1.90$ and $\beta = 2.10$ showing the equivalence of the two procedures in evaluating the average action density. Following Ref. [7] we define $\tau(n_c)$ as the gradient flow time $\tau$ for which the average action density changes by the same amount as when $n_c$ cooling steps are performed. This function is evaluated by interpolating between the discrete gradient flow time steps with cubic splines. In Fig. 7 we report the function $\tau(n_c)$ for the three different actions, for our three different ensembles. We observe that for each action used the results are in agreement for the three ensembles giving the first indication that the equivalence between the gradient flow and cooling has a well-defined continuum limit. In addition to the functions $\tau(n_c)$ we also plot the lines $\tau = n_c/3$, $\tau = n_c/4.25$ and $\tau = n_c/7.965$ for the Wilson, Symanzik tree-level and Iwasaki actions, respectively. Obviously these linear functions provide good approximations of $\tau(n_c)$ for each choice of action even for the ranges of $n_c$ where equivalence in Fig. 6 does not hold. Since the average action plays the role of a common scale between the two procedures and $\tau(n_c)$ has such a good agreement with the perturbative lines, there is little doubt that the approximation $\tau(n_c) = n_c/(3 - 15c_1)$ provides an adequate rescaling between $n_c$ and $\tau$ with finite lattice spacing corrections playing an insignificant role.

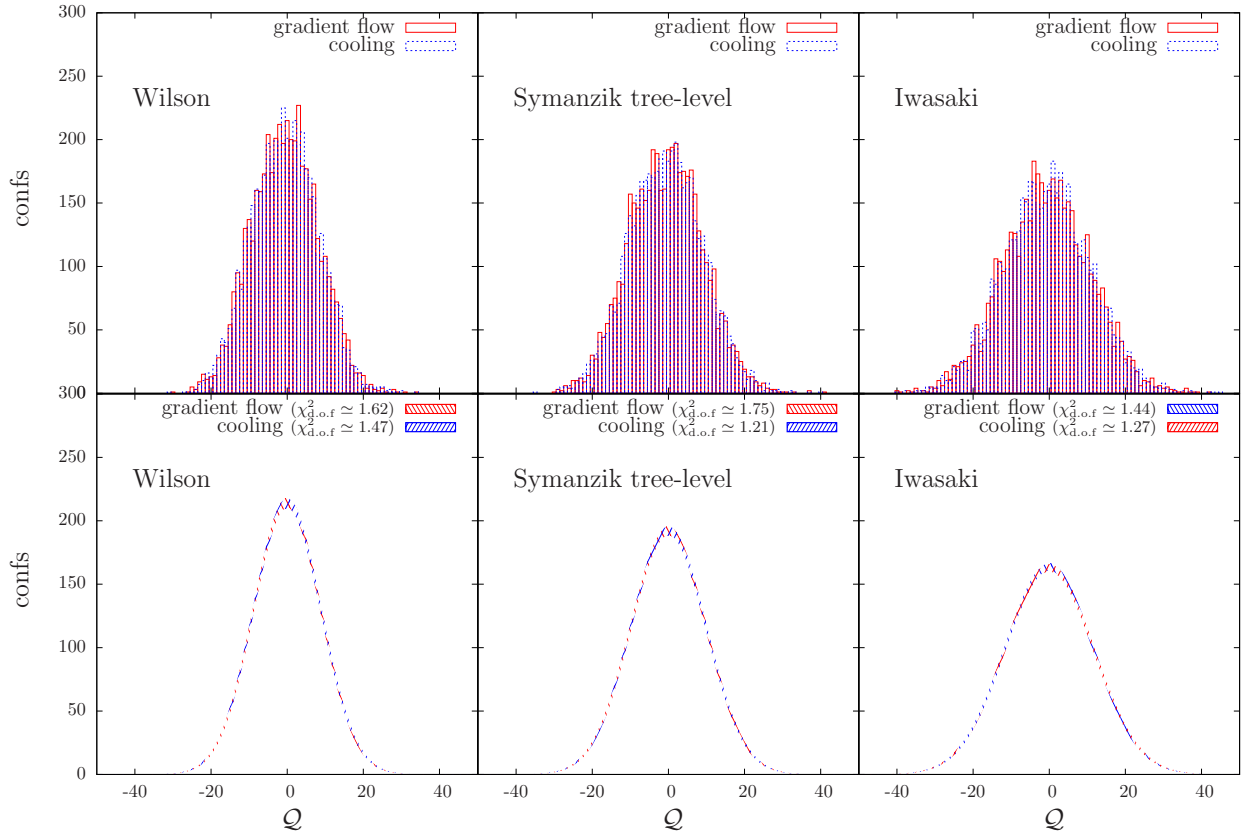

Figure 5: The distribution (first row) of the topological charge for $\beta = 1.95$ and the accosiated gaussian fit (second row). In blue we present the distribution obtained via cooling at $n_c = 50$ and in red the distribution obtained via gradient flow at $\tau = 16.7$, $\tau = 11.8$ and $\tau = 6.3$ for Wilson, Symanzik tree-level improved and Iwasaki actions, respectively.

### 5.3 Topological Susceptibility

In this section we examine results on the topological susceptibility defined as

$$\chi = \frac{\langle \mathcal{Q}^2 \rangle}{a^4 V}. \tag{45}$$

The topological susceptibility has been investigated extensively using several techniques such as smearing and cooling [22, 28, 32] and recently determinations of $\chi$ make use of the gradient flow [33] as well as the spectral projectors method [34, 35]. The question we would like to address here is not the detailed determination of the topological susceptibility, which will be the subject of another followup paper, but rather its use as a comparison between cooling and gradient flow for the three actions considered in the previous sections. In Fig. 8 we show $r_0\chi^{1/4}$ as a function of the number of cooling steps and the gradient flow time rescaled by the corresponding perturbative factor for $\beta = 1.95$. We do so for the three different actions used in the cooling and gradient flow procedure, namely the Wilson, the Symanzik tree-level improved and the Iwasaki actions, for the three lattice definitions of the topological charge density; to reveal the associated correspondence we collect the results for both procedures in the same plot. We observe that for a given action after a few cooling steps $n_c \leq 10$ or the equivalent gradient flow time $\tau = n_c/(3 - 15c_1)$ the susceptibilities computed using the plaquette or the clover definition of the topological charge density are almost indistinguishable. In general, such an agreement is not expected at finite lattice spacings and one might see deviations for very large statistics. However, for our current statistical accuracy both definitions give the same results and we thus only considered the susceptibility based on the clover definition of the topological charge. The results in Fig. 8 also show the very good agreement between cooling and the gradient flow for the topological susceptibilities obtained using the same definition for the topological charge density and the same action. As a matter of fact for even a very small number of cooling steps i.e. $n_c \sim 5$ and the corresponding gradient flow time $\tau \sim 5/(3 - 15c_1)$ the two values of the topological susceptibilities become the same. For larger number of cooling steps and the associated gradient flow times the two topological susceptibilities become almost indistinguishable. Thus, the perturbative matching between the two smoothers $\tau \simeq n_c/(3 - 15c_1)$ is confirmed as far as results on the topological susceptibility are concerned.

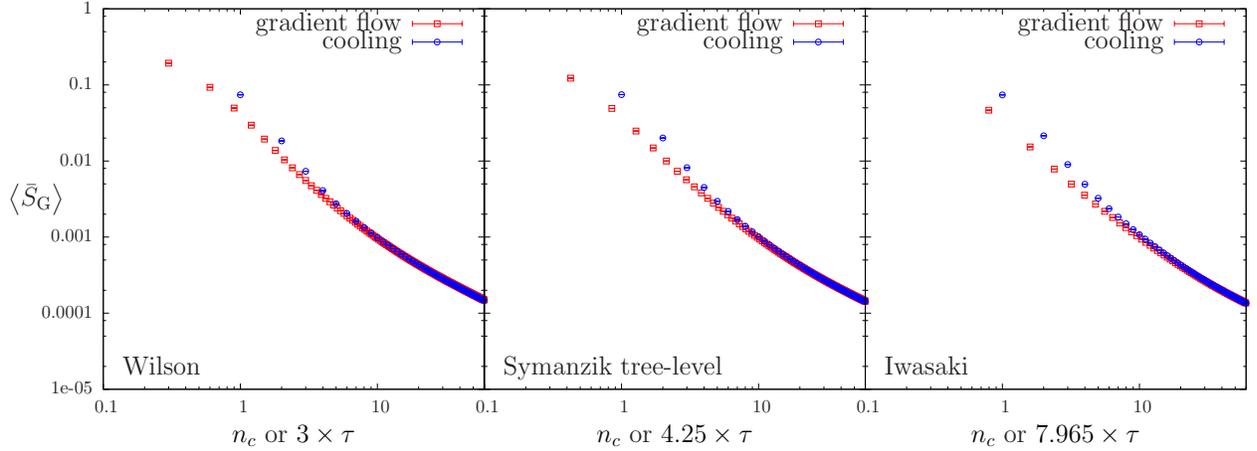

Figure 6: The average action density $\langle S_G \rangle$ as a function of the cooling step $n_c$ and the corresponding gradient flow time $n_c/(3 - 15c_1)$ for $\beta = 1.95$ and the three smoothing actions Wilson, Symanzik tree-level (tr.l) and Iwasaki.

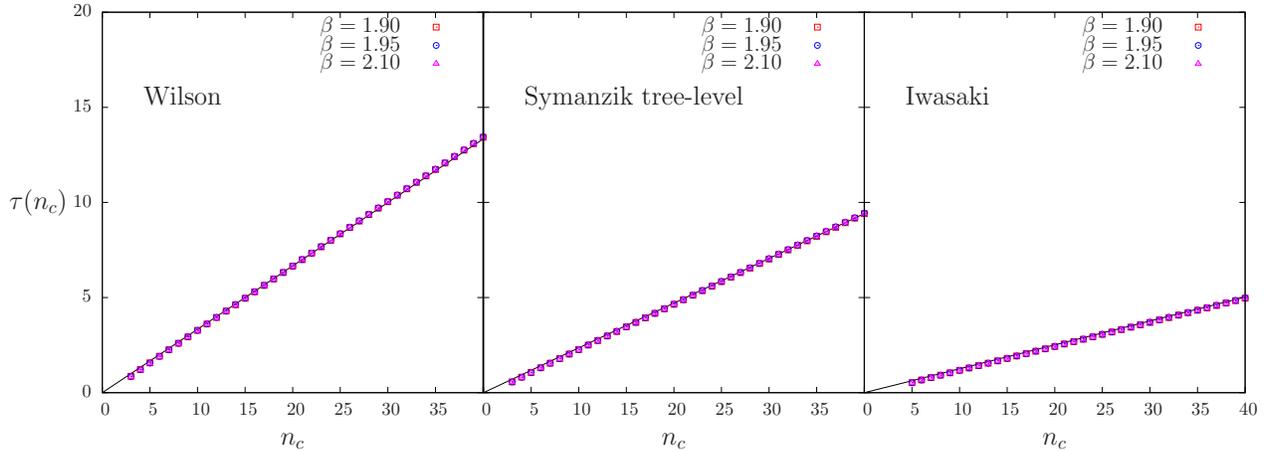

Figure 7: The behavior of $\tau(n_c)$ as a function of $n_c$ for Wilson, Symanzik tree-level improved and Iwasaki smoothing actions. The lines corresponds to $\tau = n_c/3$, $\tau = n_c/4.25$ and $\tau = n_c/7.965$ for Wilson, Symanzik tree-level improved and Iwasaki actions, respectively.

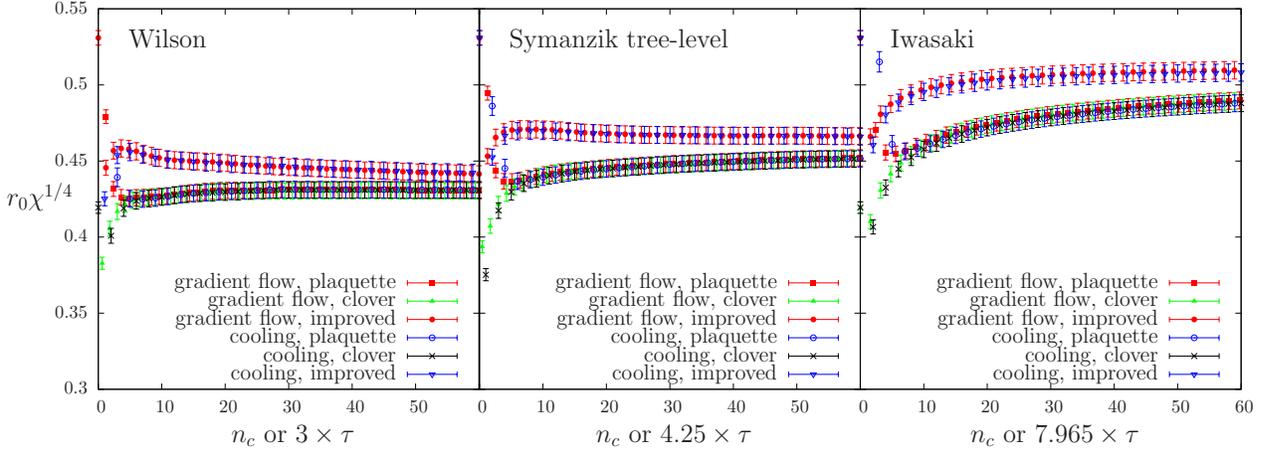

Figure 8: The topological susceptibility $r_0\chi^4$ units of $r_0$ computed using the three different definitions of the topological charge density, namely the plaquette, the clover and the improved definition, as a function of the cooling step and the associated gradient flow time. From left to right we show results for the Wilson, Symanzik tree-level improved and Iwasaki actions. The results when using the plaquette definition coincide with those obtained using the clover definition.

In Fig. 9 we present the topological susceptibility $r_0\chi^{1/4}$ as a function of the average action density defined as the common scale for cooling and the gradient flow. The susceptibility $\chi$ has been extracted for the clover and improved definitions of the topological charge density and computed using the ensembles with $\beta = 1.95$ and $\beta = 2.10$ for our three actions. We observe that for all three actions and for both definitions of the topological charge density as well as for $n_c \geq 2$ we obtain very good agreement. For our most accurate calculation using the $\beta = 1.95$ ensemble, results on $\chi$ obtained using cooling and gradient flow are in excellent agreement, but differ for the clover and improved definitions of $\chi$. Complementarily, for our finest lattice spacing ensemble with $\beta = 2.10$, we observe that the topological susceptibilities for the clover and improved definitions of the topological charge density become closer for $n_c \simeq 6, 10, 20$ for Wilson, Symanzik tree-level improved and Iwasaki smoothing actions, respectively. This is in accordance with the fact that the topological susceptibility based on the two different definitions of the topological charge density is expected to become the same towards the continuum limit.

Returning to Fig. 8 one can see that there is a plateau for the topological susceptibility as a function of the smoothing scale when the clover/plaquette definitions are used for the topological charge density which sets in when the Wilson action is used for $n_c \sim 40$. A plateau is also observed for the improved definition if the Symanzik tree-level improved action is used for $n_c \sim 40$. On the contrary, when using the Iwasaki action, the susceptibility increases with $n_c$ (or equivalently with $\tau$). This means that $n_c$ is not large enough for the Iwasaki action.

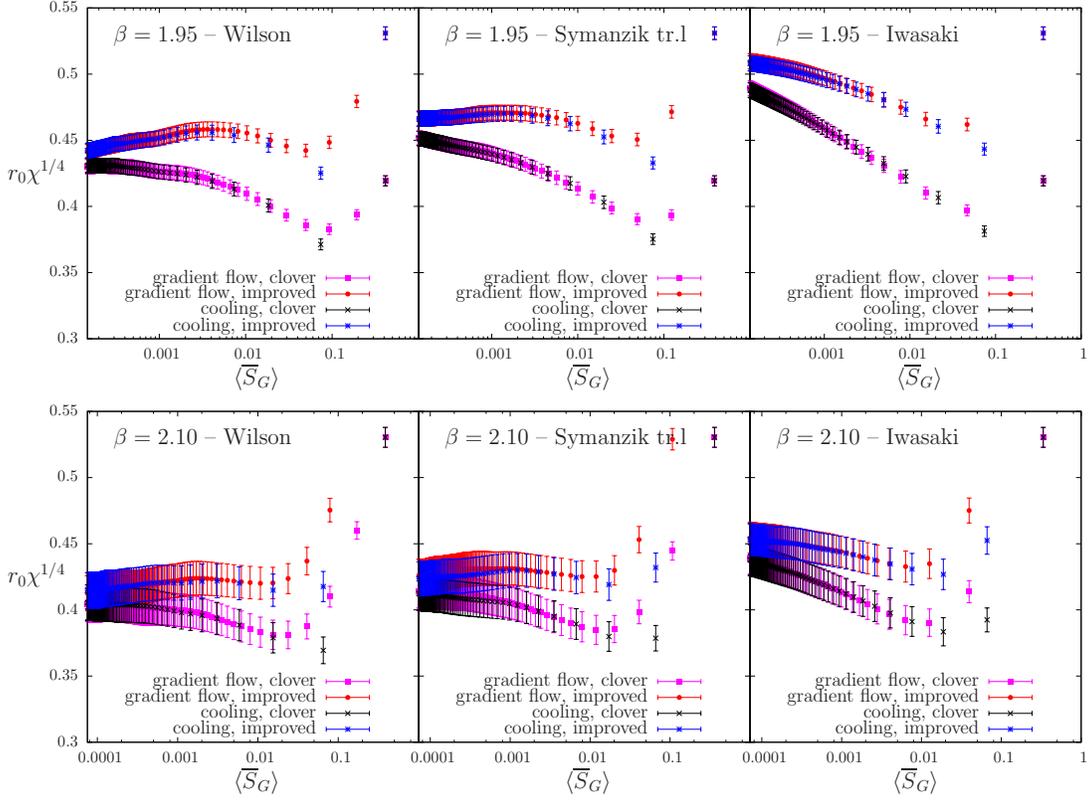

Figure 9: The susceptibility as a function of the average action density $\langle \overline{S}_G \rangle$ for $\beta = 1.95$ (top) and $\beta = 2.10$ (bottom) ensembles and the three actions.

### 5.4 Correlation Coefficient

In the previous sections we showed that cooling and gradient flow provide results, which are equivalent for the average action density and the topological susceptibility under the perturbative rescaling of Eq. (40). In this section, we examine the linear correlation coefficient for these two procedures, defined as

$$c_{\mathcal{Q}_1(n_c),\mathcal{Q}_2(\tau)} = \frac{\left\langle \left(\mathcal{Q}_1 - \overline{\mathcal{Q}}_1\right)\left(\mathcal{Q}_2 - \overline{\mathcal{Q}}_2\right)\right\rangle}{\sqrt{\left\langle \left(\mathcal{Q}_1 - \overline{\mathcal{Q}}_1\right)^2\right\rangle \left\langle \left(\mathcal{Q}_2 - \overline{\mathcal{Q}}_2\right)^2\right\rangle}}, \qquad (46)$$

where $\{\mathcal{Q}_1(n_c)\}$ and $\{\mathcal{Q}_2(n_c)\}$ are the two sets of values of the topological charge obtained using cooling at $n_c$ and gradient flow at $\tau$ respectively on the same gauge configurations. This implies that $c_{\mathcal{Q}_1(n_c),\mathcal{Q}_2(\tau)}$ is a matrix of size $n_c \times \tau$. The question we would like to answer in this section is the level of correlation between sets of topological charges extracted via cooling and gradient flow using the same action. For this discussion we employ the topological charge using the improved definition. The results for the other two definitions are similar. In Fig. 10 we represent results for the correlation coefficient using the three actions for our three ensembles.

We show the diagonal elements (for $n_c = (3-15c_1)\tau$) of the correlation coefficient matrix $c_{\mathcal{Q}_1(n_c),\mathcal{Q}_2(\tau)}$ when $n_c$ and $\tau$ are matched with the perturbative expression Eq. (40). When the Wilson action is used, we observe that for $n_c > 1$ and as we increase $n_c$ the coefficient drops till it reaches a nearly stable value ($n_c > 10 - 20$). This value is approximately $\sim 93.5\%$ for $\beta = 1.90$, $\sim 95\%$ for $\beta = 1.95$ and $\sim 98\%$ for $\beta = 2.10$. Clearly, as $a \to 0$ the correlation coeffient approaches unity. This indicates that the correspondence between cooling and gradient flow has a well–defined continuum limit. A similar behavior is observed when the Symanzik tree-level improved action is used obtaining $\sim$93% for $\beta = 1.90$, $\sim$95% for $\beta = 1.95$ and $\sim$97.5% for $\beta = 2.10$. Finally and likewise when the Iwasaki smoothing action is used the level of correlation is $\sim$92.5% for $\beta = 1.90$, $\sim$94% for $\beta = 1.95$ and $\sim$96.5% for $\beta = 2.10$.

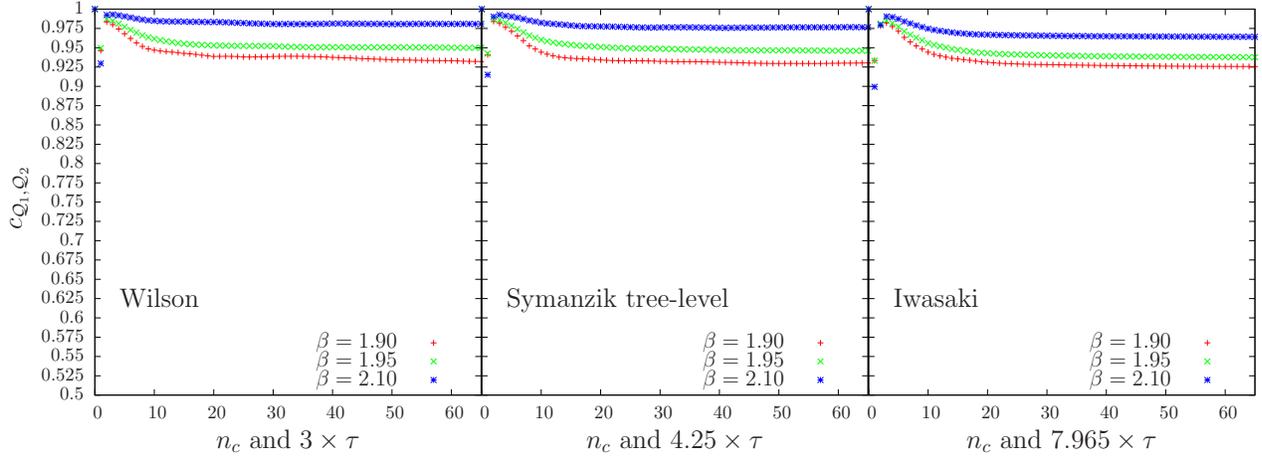

Figure 10: The diagonal elements of the correlation coefficient $c_{Q_1(n_c),Q_2(\tau)}$ defined in Eq. (46) for the Wilson, Symanzik tree-level improved and Iwasaki actions respectively for $\beta = 1.90$, $\beta = 1.95$ and $\beta = 2.10$. We consider the topological charge extracted using the improved definition.

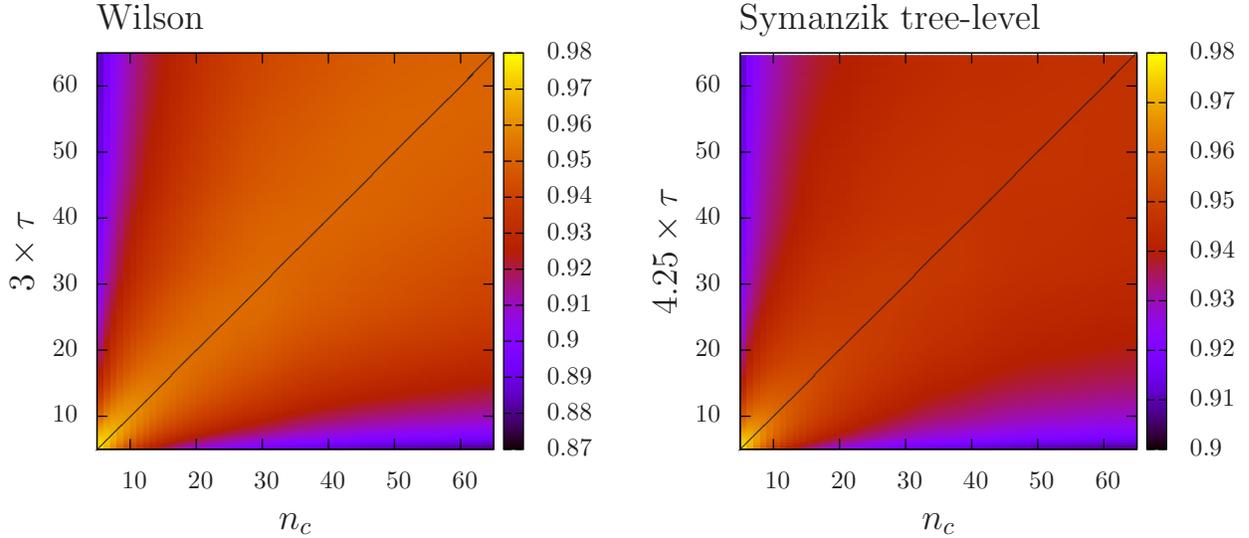

Figure 11: The correlation coefficient matrices $c_{Q_1(n_c)Q_2(\tau)}$ for the Wilson and Symanzik tree-level improved action. We consider topological charge extracted for $\beta = 1.95$ and the improved definition of the topological charge density.

In Fig. 11 we provide density plots for the full correlation coefficient matrix $c_{Q_1(n_c)Q_2(\tau)}$ for the Wilson and Symanzik tree-level improved actions obtained when the improved topological charge is employed for $\beta = 1.95$. When excluding the very first cooling steps (e.g. $n_c \lesssim 10$), the matrix $c_{Q_1 Q_2}$ appears to be nearly diagonal with the diagonal line denoting the equation $n_c = (3 - 15c_1)\tau$. This behavior is more pronounced for the case of the Wilson and Symanzik tree-level improved actions. Thus these results corroborate the fact that cooling the gauge configurations with $n_c$ steps has almost the same effect as evolving these configurations via gradient flow for $\tau = n_c/(3 - 15c_1)$. We expect that at the continuum limit the corresponding distributions become perfectly diagonal with the maximum along the diagonal and corresponding to a correlation coefficient of 100%.

# 6 Conclusions

In this article we provide a comparison of the results on observables such as the topological charge and the susceptibility obtained using gradient flow or cooling. It extends the analysis of Ref. [7] to include gauge actions with rectangular terms. The comparison is realized both analytically in perturbation theory and numerically. For our analytic analysis we follow the perturbative treatment of Ref. [7], which was performed for the Wilson action and we show how to generalize it to Symanzik improved actions with rectangular parts. More specifically, we derive the corresponding relation between the continuous gradient flow time $\tau$ and the number of the discrete cooling steps $n_c$ by expanding the flow steps perturbatively including terms up to $\mathcal{O}(a)$. The relation we obtain is $\tau \simeq n_c/(3 - 15c_1)$ where $c_1$ is the coefficient which multiplies the rectangular term in the gauge action. This becomes exact as $a \to 0$ and does not depend on the details of the gauge group; although this is derived for $SU(2)$ the generalization to $SU(3)$ is straight forward. For the numerical results we use configurations produced with $N_f = 2+1+1$ twisted mass fermions and the Iwasaki gauge action. Although strictly speaking the relation we derived is valid only as $a \to 0$, we confirm numerically that the action density, used as a common scale, coincides for both procedures.

By investigating the time histories of the topological charge we observe that these behave in the same manner for both smoothing procedures indicating equivalence between them. The histograms of the topological charge distributions for fixed $n_c$ and $\tau \sim n_c/(3 - 15n_c)$ are almost the same for both smoothers and approximately Gaussian having the same width. This already suggests an equivalence for the topological susceptibility, which is confirmed by calculating the topological susceptibility $\chi$ for all three lattice definitions of the topological charge density as a function of the smoothing scale and the average action for both smoothers. This enables us to demonstrate that after a very few cooling steps $n_c \sim 2$ the topological susceptibility for gradient flow and cooling become equivalent; this holds for all tested smoothing actions and all lattice definitions of the topological charge.

Finally we look at the correlation coefficient, which can be used to reveal similarities between the different definitions of the topological charge. We observe maximum correlation for gauge configurations that have been smoothed via gradient flow or cooling according to the relation $\tau \simeq n_c/(3 - 15c_1)$. In addition, we observe that after a few cooling steps the correlation coefficient becomes stable with increasing value towards the unity as we approach the continuum limit (decreasing the lattice spacing). For instance already for our finest lattice with $\beta = 2.10$ the correlation coefficient is $\sim 98\%$, $\sim 97.5\%$ and $\sim 96.5\%$ when smoothing with Wilson, Symanzik tree-level improved and Iwasaki action, respectively.

The main conclusion of this study is that one can use cooling or the gradient flow in order to extract the topological properties of configurations smoothed with gauge actions, which include square and rectangular terms. This equivalence is manifested by using the relation $\tau = n_c/(3 - 15c_1)$ derived in perturbation theory. In practice, this means that one may opt to use cooling to extract the topological charge $\mathcal{Q}$. An approximate comparison between the gradient flow time $\tau$ with integration step $\epsilon = 0.01$ and cooling step $n_c$ for an action, which includes rectangular terms gives cpu_time($\tau = 1$)/cpu_time($n_c = 1$) $\simeq 160$. Hence, for the Symanzik tree-level improved action, gradient flow is slower than cooling by a factor of $\sim 38$ while for the Iwasaki action by $\sim 20$. These estimates depend on the integrator used for the gradient flow and the integration step $\epsilon$. The speed-up cooling gives in comparison to gradient flow is $O(10)$ and this could decrease the computational cost by the same factor in investigations where one is mainly interested in the topological susceptibility and where a large number of configurations is required.

## Acknowledgments

We would like to thank all members of ETMC for the most enjoyable collaboration. Numerical calculations have used HPC resources from John von Neumann-Institute for Computing on the JUQUEEN and JUROPA systems at the research center in Jülich. Additional computational resources were provided by the Cy-Tera machine at The Cyprus Institute funded by the Cyprus Research Promotion Foundation (RPF), ΝΕΑΥΠΟΔΟΜΗ/ΣΤΡΑΤΗ/0308/31. A.A is indebted to K. Cichy, B. Lucini, H. Panagopoulos and M. Teper for enlighting discussions on topology, and A. Ramos for usefull discussions on the implementation of the gradient flow. A. A. acknowledges the hospitality of the Cyprus Institute where part of this work was carried out.